\pdfoutput=1
\documentclass{JINST}
\usepackage{graphicx}
\usepackage{amssymb}

\usepackage{amsmath}
\usepackage{xspace}
\usepackage{booktabs}
\usepackage{url}
\def \fe55 {$^{55}$Fe\xspace}
\def \cstwo {CS$_{2}$\xspace}

\newcommand{\driftII}[1]{DRIFT\nobreakdash-II#1} 

\DeclareFontFamily{U}{euc}{}
\DeclareFontShape{U}{euc}{m}{n}{<-6>eurm5<6-8>eurm7<8->eurm10}{}%
\DeclareSymbolFont{AMSc}{U}{euc}{m}{n} 
\DeclareMathSymbol{\umu}{\mathord}{AMSc}{"16} 

\def \microsec {\ensuremath{\umu}s\xspace}

\title{Low Energy Electron and Nuclear Recoil Thresholds in the DRIFT-II Negative Ion TPC for Dark Matter Searches}
\author{S.\ Burgos$^a$, E.\ Daw$^b$, J.\ Forbes$^a$, C.\ Ghag$^c$, M.\ Gold$^d$, C.\ Hagemann$^d$, V.A.\ Kudryavtsev$^b$, T.B.\ Lawson$^b$, D.\ Loomba$^b$, P.\ Majewski$^b$, D.\ Muna$^b$\thanks{Corresponding author.}, A.\ St.J.\ Murphy$^c$, S.M.\ Paling$^b$, A.\ Petkov$^a$, S.J.S.\ Plank$^c$, M.\ Robinson$^b$, N.\ Sanghi$^d$, D.P.\ Snowden-Ifft$^a$, N.J.C.\ Spooner$^b$, J.\ Turk$^d$, and E.\ Tziaferi$^b$\\
\llap{$^a$}Department of Physics, Occidental College,\\
	Los Angeles, CA 90041, USA\\
\llap{$^b$}Department of Physics and Astronomy, University of Sheffield\\
	Sheffield S3 7RH, UK\\
\llap{$^c$}School of Physics and Astronomy, University of Edinburgh\\
	Edinburgh EH9 3JZ, UK\\
\llap{$^d$}Department of Physics and Astronomy, University of New Mexico,\\
	Albuquerque, NM 87131, USA\\
Email: \email{demitri.muna@nyu.edu}}

\abstract{Understanding the ability to measure and discriminate particle events at the lowest possible energy is an essential requirement in developing new experiments to search for weakly interacting massive particle (WIMP) dark matter. In this paper we detail an assessment of the potential sensitivity below 10~keV in the 1~m$^{3}$ DRIFT-II directionally sensitive, low pressure, negative ion time projection chamber (NITPC), based on event-by-event track reconstruction and calorimetry in the multiwire proportional chamber (MWPC) readout. By application of a digital smoothing polynomial it is shown that the detector is sensitive to sulfur and carbon recoils down to 2.9 and 1.9~keV respectively, and 1.2~keV for electron induced events. The energy sensitivity is demonstrated through the 5.9~keV gamma spectrum of $^{55}$Fe, where the energy resolution is sufficient to identify the escape peak. The effect a lower energy sensitivity on the WIMP exclusion limit is demonstrated. In addition to recoil direction reconstruction for WIMP searches this sensitivity suggests new prospects for applications also in KK axion searches.}

\keywords{Dark matter; NITPC; MWPC}

\begin{document}

\section{Introduction}
\label{section:introduction}

The DRIFT dark matter experiment \cite{Alner:2005aa,drift-alpha-paper} is designed to identify weakly interacting massive particles (WIMPs) \cite{2004ARNPS..54..315G}, the leading candidate for dark matter, not just by searching for the WIMP-induced nuclear recoils expected but also through measurement of their direction. This is feasible in DRIFT through use of low pressure negative ion time projector chamber technology (NITPC) \cite{nitpc-2} in which recoil tracks generated within a gas volume are drifted in an electric field and read out by two back-to-back multi-wire proportional chambers (MWPC) \cite{first-drift-iia-analysis-paper}. The signature of a positive detection of dark matter is then a distribution of nuclear recoil directions that is correlated to the Earth's motion through the galactic halo. This is known to be a highly robust identification signature for dark matter uncluttered by too many model-dependent assumptions \cite{2005PhRvD..71j3507M}. The basic directional sensitivity and operational capability of the technology for WIMP searches has been investigated in \cite{d2-directional-signature,first-drift-iia-analysis-paper}. In this paper we present work to determine the potential energy thresholds of the basic detector technology and hence the degree to which low energy (sub 10~keV) electron and nuclear recoil events can be recorded. 

Determining the energy threshold for recording nuclear recoils is a key parameter required to assess the ultimate capability of any experiment to standard WIMPs. However, further motivation is provided here by the potential ability of DRIFT to also track low energy electron and gamma events. This opens the possibility of applying DRIFT to searches for KK axions or decay of axion-like particles to two gammas \cite{2005APh....23..287M}. The ability to identify coincident EM event pairs well separated in space due to the use of a low pressure gas allows for powerful background rejection for such searches and hence unique sensitivity, provided a gamma threshold below 5~keV can be achieved \cite{2005APh....23..287M}. In this situation there is potential also to identify and reconstruct multi-prong events involving coincident production of both nuclear recoils and electrons or gammas. Such events have been suggested as an explanation for the DAMA dark matter signal. DAMA observe an annual modulation of event rate in their 250 kg NaI detector array for events $<$6 keV, with phase consistent with that expected for WIMPs due to the Earth's solar orbit.

To study these threshold issues use was made here of data from the DRIFT-IIb detector taken using \fe55 low energy (5.9 keV) gamma sources, during operation underground at the Boulby underground laboratory. The DRIFT-IIb detector is second in a line of three essentially identical DRIFT NITPC detectors built for directional studies. Details of the design are given in \cite{Alner:2005aa}. Briefly, the detector consists of a 1.5~m$^{3}$ stainless steel vacuum vessel filled with gaseous \cstwo held at 40~Torr (0.053~atm). A central cathode plane divides the 0.76~m$^{3}$ fiducial volume into two TPCs, each with a drift length of 50~cm. Ionisation tracks created in the TPC drift away from the central cathode towards one of two MWPCs. Each MWPC consists of two planes (cathode and anode) of 512~wires each, orientated perpendicularly to provide $x$ and $y$ co-ordinate information. Wires are grouped down into eight channels per plane. The outer 26~wires on each edge of the planes serve as a veto readout to identify events that occur at the edges of the fiducial volume or events that originated from outside the TPC (e.g. alpha particles). Each MWPC has an \fe55 source located 15~cm behind the wire plane which is exposed to the chamber via a retractable shield during energy calibrations.

\section{$^{55}$Fe X-rays in \cstwo}

The DRIFT-II detector measures the number of ions pairs (NIPs) generated of a track produced in the fiducial volume from an ionising source (nuclear recoils, alpha particles, etc.). A calibration must be performed to convert the NIPs value into an energy $E$ using the relation $E/W= N$,
where $W$ is a value that depends on the ionising source, the target gas, and the incident energy, and $N$ is the mean number of NIPs generated \cite{ionisation-energy} by a track. A source with a monochromatic energy is used to produce tracks in the chamber, and the number of NIPs measured provides a calibration point to the known energy.

\fe55 is commonly used for energy calibration in MWPCs \cite{charpak-sauli-review,2007NJPh....9..171A}, and a 100~$\umu$Ci source is used in the DRIFT-II experiment. The dominant \fe55 decay produces a 5.895~keV X-ray, and the probability of an interaction between this X-ray and either a carbon or sulphur atom (the components of the \cstwo target mass) is determined from the mean free path, $\lambda$. The density of 40~Torr \cstwo is 1.67~$\cdot$~10$^{-4}$~g/cm$^{3}$, and with the photoelectric absorption coefficient $\mu$ for carbon of 11.1~cm$^{2}$/g gives a mean free path of 539.5~cm (compared to a maximum drift distance of 50~cm). The value of $\mu$ for sulphur is 220.0~cm$^{2}$/g, giving a mean free path of 27.2~cm. As the mean free path for a carbon interaction is large compared to the size of the detector and there are twice as many sulphur atoms, it is reasonable to assume all X-ray events are gamma-sulphur interactions. Table~\ref{table:mean-free-paths} lists the mean free path for different energies of photons in 40~Torr \cstwo by carbon and sulphur. (Photoelectric absorption values obtained from the National Institute of Standards and Technology \cite{nist-xcom}.)

\begin{table}[t]
\begin{center}
\begin{tabular}{cr@{.}lr@{.}lr@{.}l}
\toprule
	Process &
	\multicolumn{2}{c}{$\gamma$ Energy} &
	\multicolumn{2}{c}{$\lambda$ in C} &
	\multicolumn{2}{c}{$\lambda$ in S}\\
\midrule
\fe55 X-ray & 5&895 keV & 539&46 cm & 27&22 cm\\
S fluorescence & 2&241 keV & 27&85 cm & 21&31 cm\\
\bottomrule

\end{tabular}

\caption[Mean Free Path Interaction Lengths]{Mean free interaction path $\lambda$ for carbon and sulfur nuclei in 40~Torr \cstwo.}
\label{table:mean-free-paths}

\end{center}
\end{table}

Two processes (Auger emission and fluorescence) primarily contribute to the \fe55 energy spectrum producing three energy signatures. The excitation of the sulphur atom produces 3.426~keV photoelectrons and Auger emission electrons, together creating a single ionisation track. This results in the full absorption peak located at 5.895~keV. Auger emission is the dominant mechanism with a yield in sulphur of 92.4\%, while the fluorescence yield is 7.6\% \cite{nist-xcom}. In the latter case, the loss of the fluorescence X-ray produces an escape peak with an energy of the original photoelectron. DRIFT differs from most simple proportional counters in that it has a very large fiducial volume, as such, the mean free interaction path of the fluorescence X-ray in sulphur of 21.31~cm (Table~\ref{table:mean-free-paths}) means that a percentage will produce another track away from (and thus uncorrelated to) the initial event but still within the fiducial volume. These events will produce a fluorescence capture peak with an energy at 2.241~keV, the energy of the fluorescence X-ray.

\begin{figure}[tp]
\begin{center}
\begin{tabular}{c}

	\includegraphics[width=1.0\textwidth]{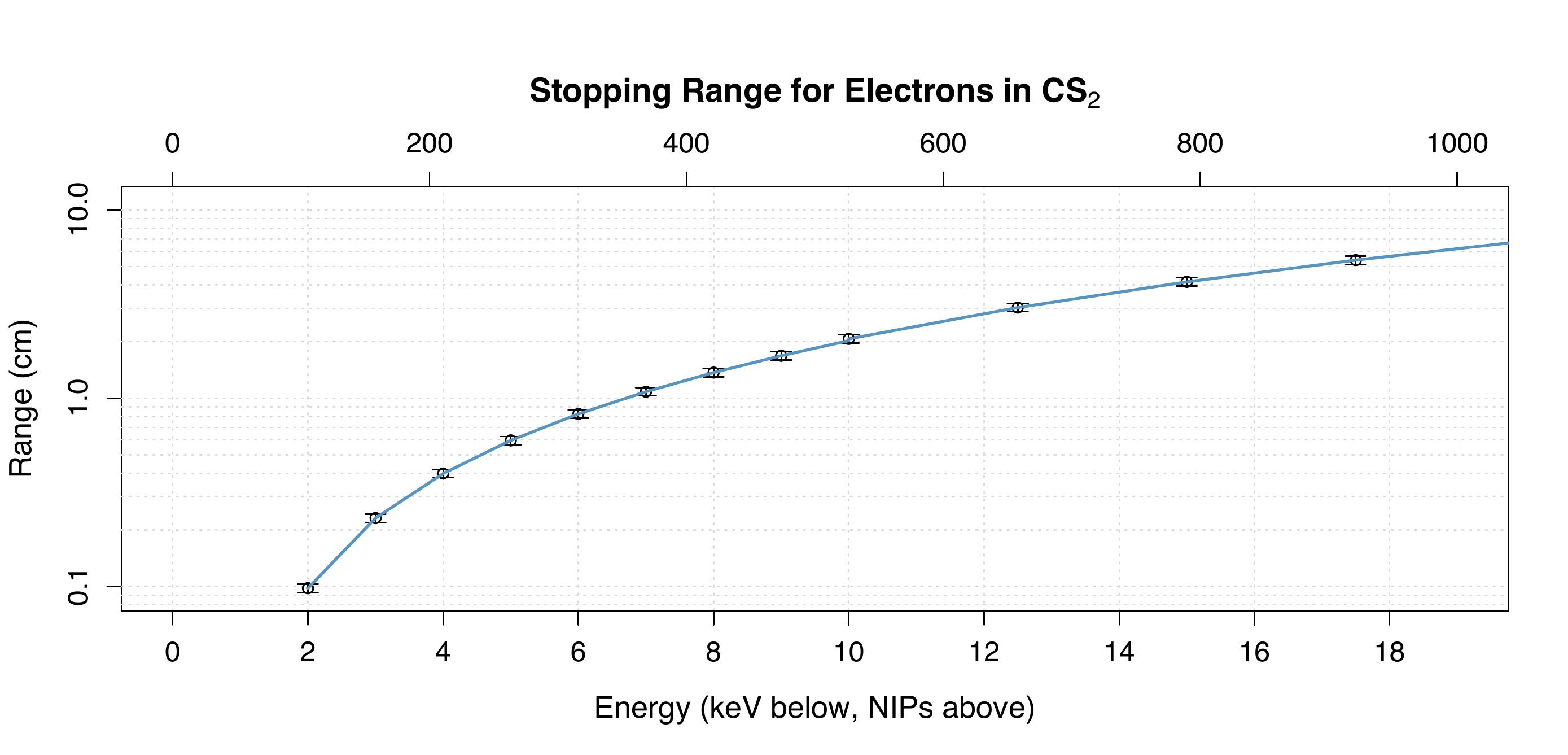}

\end{tabular}
\end{center}
\caption[Electron Stopping Range in 40 Torr \cstwo]{The stopping range of electrons in 40~Torr \cstwo (data from \cite{estar-nist}).}

\label{figure:electron-stopping-range}
\end{figure}

\begin{figure}[tp]
\begin{center}
\begin{tabular}{c}

\includegraphics[width=0.8\textwidth]{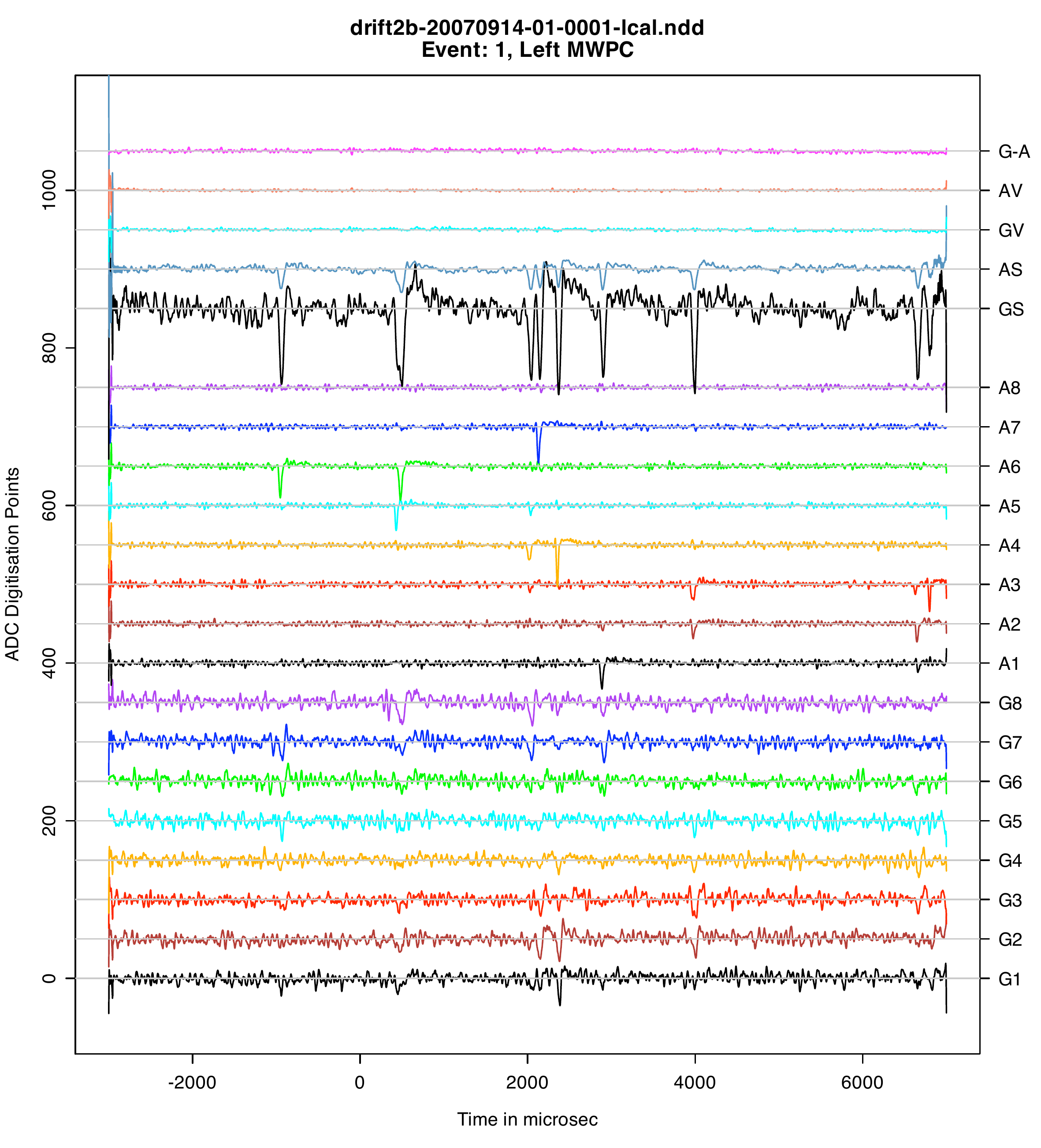}\\

\end{tabular}
\end{center}
\caption[\fe55 Event]{An event record containing at least ten \fe55 events. Most events primarily fall on a single wire (e.g.\ $t=-1000$~\microsec, A6), while others appear to be split between two (e.g.\ $t=4000$~\microsec, A2 and A3). The analysis software attempts to reconstruct simultaneous events falling on two adjacent wires into a single track. Anode MWPC wires are labelled A1 through A8; cathode (or grid) MWPC wires are similarly G1--G8. Above these are the sum of all anodes (AS), the sum of all cathode (GS), the anode (AV) and grid (GV) vetoes, and the difference between the vetoes (G--A). Signals at the edges of the event record are a side effect of the digital smoothing; these are excluded from analysis.}

\label{figure:fe55-event}
\end{figure}

\section{Event Selection and Digital Filtering}
Figure~\ref{figure:fe55-event} shows a typical \fe55 event, where at least ten X-ray events appear. The waveforms shown are after a well understood 55~kHz noise in the detector has been removed through a Fourier analysis. The signals on the anode wires (labelled A1--A8 in Figure~\ref{figure:fe55-event}) exhibit less noise than those on the cathode wires as the latter provide shielding from the high voltage of the central cathode; consequently, these are used for the energy calibration. The pulses selected for analysis are those that fall below a software threshold (in these plots, charge deposition results in a negative pulse). This threshold is determined separately for each channel set as four times the standard deviation of the waveform. The height is measured as the furthest point of the waveform from the baseline in units of ADC counts for each pulse, and the area is the integration between baseline crossings in units of ADC-\microsec. These are the only two parameters required for the purpose of this calibration.

Analysing events in these data is not solely a matter of counting pulses as would be the case for a simple proportional counter. The large volume coupled with the fact that tracks typically cross more than one wire requires additional reconstruction. The range of the \fe55 events in \cstwo is on the order of the 2~mm wire spacing, however, it is possible for the track, and subsequently the charge, to be split across two wires. One example of this can be seen in Figure~\ref{figure:fe55-event} at 4000~\microsec on channels A2 and A3. The events are reconstructed as tracks to account for this. Tracks that are spread across two wires account for 54\% of the events in the calibration presented here. We define times $t_{1}$ and $t_{2}$ as the left and right edges of the full width half maximum of the pulse. Two pulses are defined as being part of the same track if they are on adjacent wires and either $t_{1}$ or $t_{2}$ falls within 50~\microsec of the second pulse's $t_{1}$ or $t_{2}$. This value is chosen to minimise counting pile-up events as a single event (e.g. the events on channels A5 and A6 at $t\approx200$~\microsec in Figure~\ref{figure:fe55-event}). While this is unavoidable, the summed area of pile-up events joined as a track will typically be sufficiently above the full absorption peak so as not to contaminate the resulting spectrum.

\begin{figure}[tp]
\begin{center}
\begin{tabular}{c}
	\includegraphics[width=1.0\textwidth]{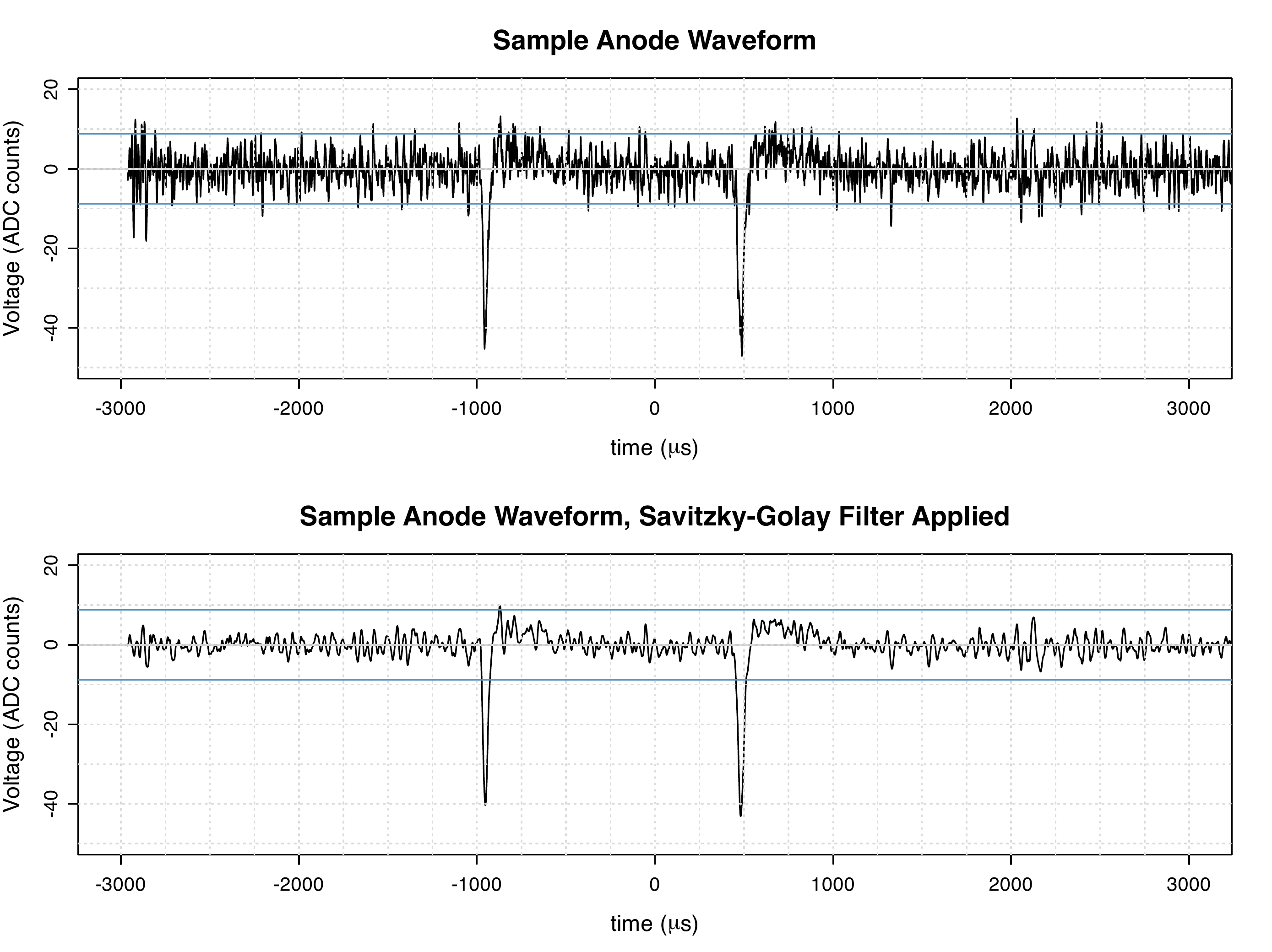}
\end{tabular}
\end{center}

\caption{Demonstration of noise filtering with the Savitzky-Golay digital filter. Above, a sample waveform from an anode channel showing two \fe55 events. Below, the same with the digital filter applied. A $4\sigma$ threshold is indicated. Thirty-three pulses trigger the software analysis above and are mostly noise, where the only two below are actual events. Both waveforms are shown after a 55~MHz noise removal via Fourier analysis.}

\label{figure:waveform-comparison}
\end{figure}

The level of noise in the signal presents a limit to how sensitive the analysis is to low energy events. The standard Savitzky-Golay algorithm \cite{savitzky-golay}, a digital smoothing polynomial filter, is applied to the data to reduce the standard deviation of the noise level by a factor of 5.0 compared to the raw data \cite{muna-thesis}. A fourth order fit is used with a 65~\microsec (65~samples) smoothing window. This filter was chosen for its ability to preserve pulse area, centre of gravity, line width, and pulse symmetry \cite{dispo-filters}. The results are shown in Figure~\ref{figure:waveform-comparison}. The threshold indicated in both plots is the 4$\sigma$ level of the noise of the smoothed waveform, demonstrating that the analysis can apply a lower threshold than the level of the noise in the raw data without constantly triggering on noise. This allows events of lower energy to be extracted that would have otherwise been buried in the noise, resulting in an increase of the sensitivity of the detector. The sharp lines that appear at the beginning and end of most of the traces in Figure~\ref{figure:fe55-event} are artefacts of this filter; as such all signals on the edges of the event record are excluded from analysis. Selecting events based on a multiplier of the noise level represents a first pass of event selection; further analyses may employ a fixed threshold above the baseline.

\section{Experimental Data}
An \fe55 calibration of each MWPC is performed every six hours during a normal experimental run, collecting 1000 triggered events in approximately two minutes. This is sufficient to measure the full absorption peak required for the calibration. The results of this paper are instead based on an extended \fe55 exposure collecting twenty thousand event records in order to collect a large statistical sample to demonstrate the potential energy sensitivity of the MWPC. Nearly seven thousand event records from these were identified as containing unwanted events (alpha particls, sparks, etc.) and removed from the analysis (see \cite{muna-thesis} for details of cuts).

\begin{figure}[tp]
\begin{center}
\begin{tabular}{c}
	\includegraphics[width=0.8\textwidth]{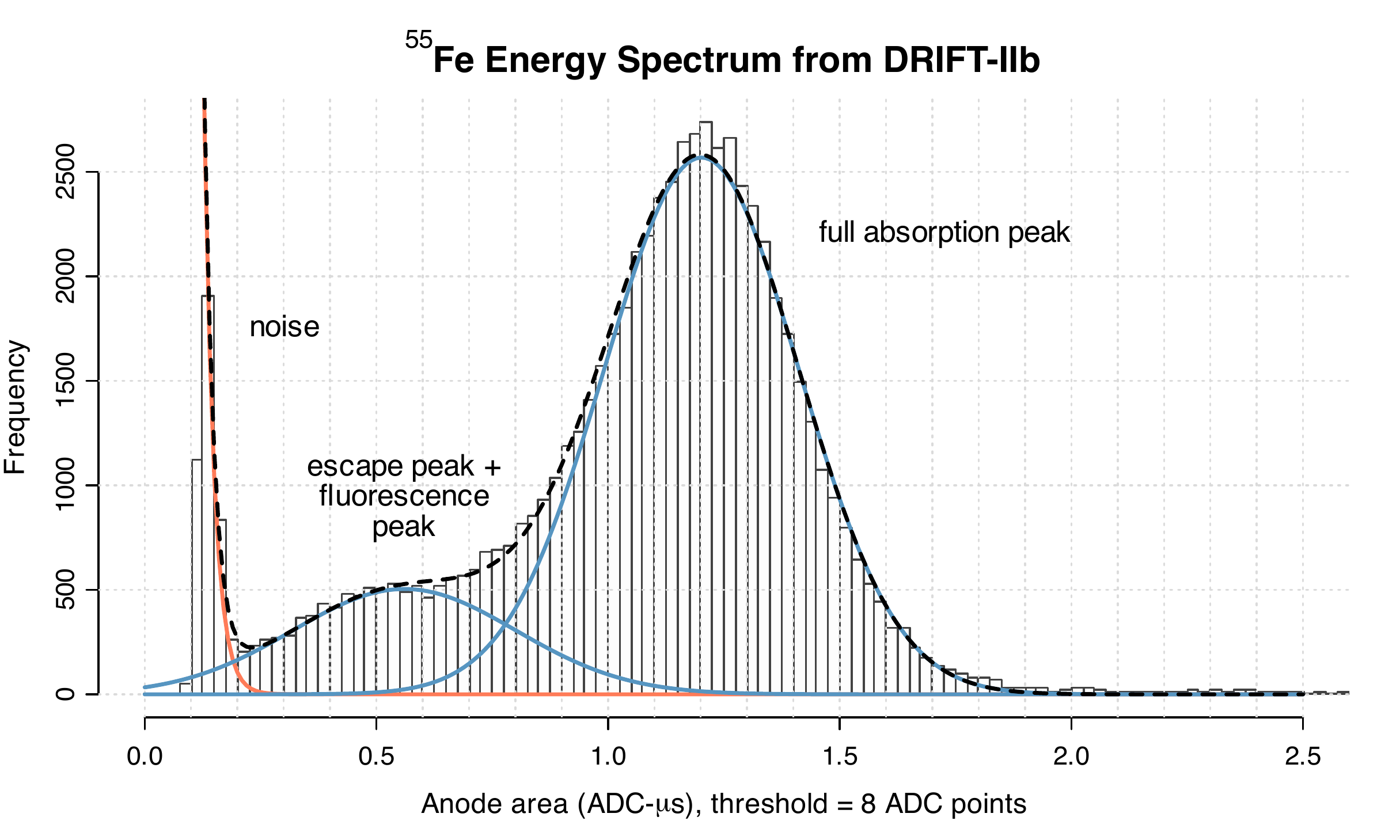}

\end{tabular}
\end{center}

\caption{An \fe55 energy spectrum from experimental data from DRIFT-IIb. Both track reconstruction and digital polynomial smoothing have been applied to the data. The data is fit to an exponential decay (noise) plus two Gaussian curves. The first is the full absorption peak, and the second is the sum of the escape and fluorescence peaks (which cannot be individually resolved).}

\label{figure:full-spectrum}
\end{figure}

The experimental energy spectrum is expected to be comprised of three components: a noise peak, the escape plus fluorescence peak, and the full absorption peak. This is modelled as an exponential function and two Gaussian functions. Figure~\ref{figure:full-spectrum} is the result of the full analysis after track reconstruction and Saitzky-Golay smoothing. Although the smaller peak is the sum of the escape peak and the fluorescence peak, these features are too close in energy to be individually resolved. The full absorption peak is clearly evident and resolved from the smaller peak to the left, which in turn is separated from the noise at the lowest energies. The presence of the noise peak indicates that the events recorded are not subject to a pulse height threshold that is too high --- the events that comprise the \fe55 features are above the noise.

A comparison of the ratio of the mean positions between the two peaks provides a validation of the results. The smaller peak consists of 2.2 and 3.4~keV events, while the full absorption peak will be 5.895~keV. The escape peak events and the fluorescence events do not occur in equal number; the number of fluorescence events counted is dependent on the detector geometry. A Monte Carlo was performed and found that the fluorescence peak event count should be 6\% of the full absorption peak (i.e.\ 79\% of the fluorescence X-rays interact within the fiducial volume and do not escape). Using the fluorescence yield of 7.6\%, this results in a theoretical ratio of 2.30 between the mean position of the full absorption peak and the smaller peak as measured by the detector. The experimental ratio is 2.15~$\pm$~0.13, showing these values to be in reasonable agreement.

\begin{figure}[tp]
\begin{center}
\begin{tabular}{cc}

\includegraphics[width=1.0\textwidth]{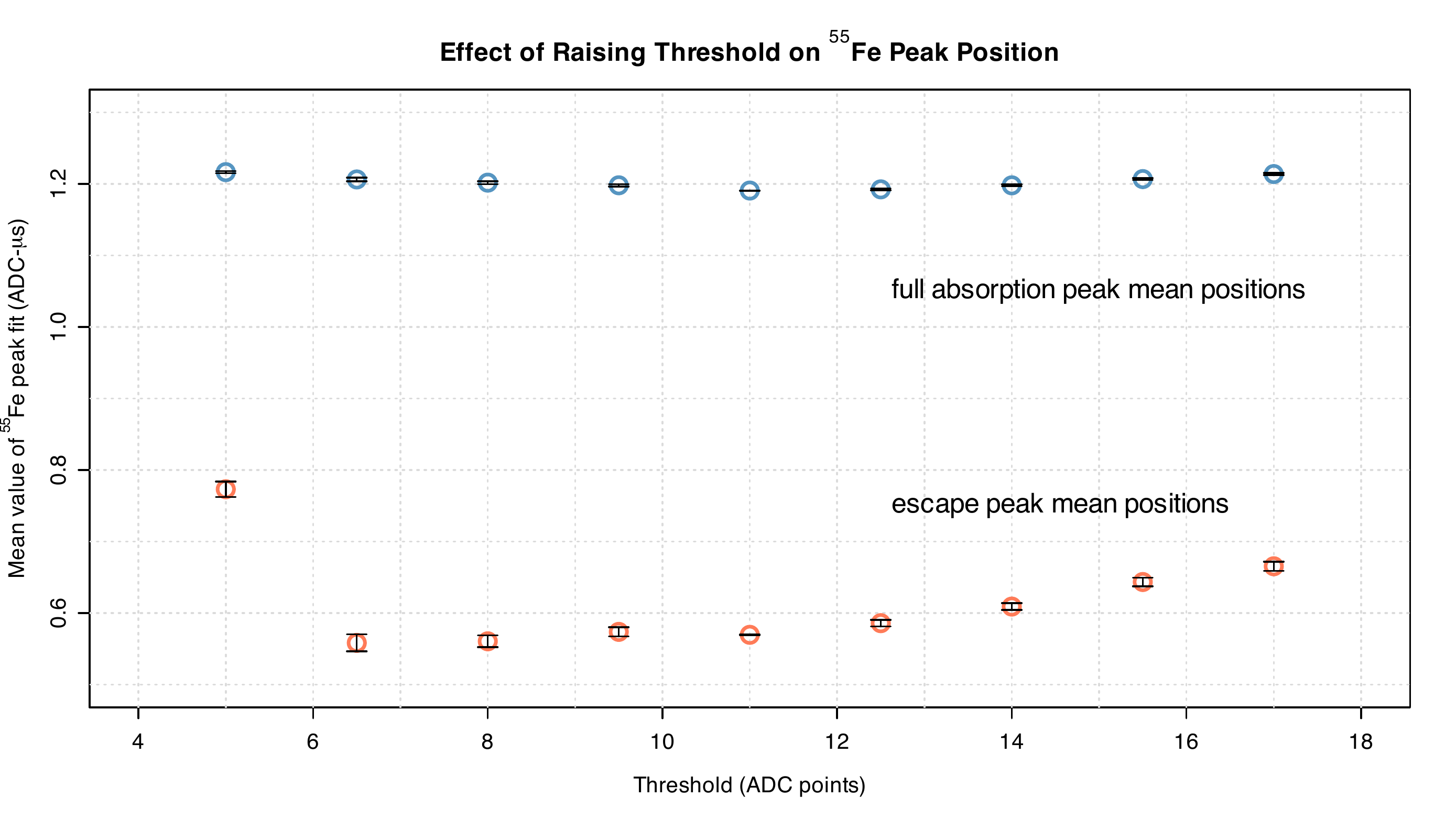}

\end{tabular}
\end{center}
\caption[\fe55 Peak Positions vs.\ Threshold]{A plot of the means of the escape and full absorption peaks with varying thresholds after the digital filter is applied.}

\label{figure:fe55-peak-vs-threshold}
\end{figure}

The software threshold can be varied in order to further ensure that the features are real; the mean positions of the peaks should not be a function of the software threshold. Figure~\ref{figure:fe55-peak-vs-threshold} shows the mean position of both the escape and full absorption peaks with fixed thresholds ranging from 5 to 17 ADC counts. (The first point is effectively applying no threshold above the software threshold as the minimum height is 5.35~ADC counts.) The full absorption peak remains constant (to within less than 2\%). The first point of the escape peak shows sensitivity to noise, but a threshold of 6.5~ADC counts effectively removes the majority of noise from the energy region. The values remain stable, but a threshold of $\sim$12~ADC counts and above begins to remove events from the escape peak and the feature begins to be lost. Not much more above this, and the feature is removed altogether. This ability to identify genuine physics features at such a low energy provides a validation of the smoothing algorithms performed on the data set (the peak-to-peak noise level in the unsmoothed data is about $\pm$10~ADC counts).

The energy resolution of the detector can be calculated from this data as $\sigma/E_{0}$ where $E_{0}$ is the mean position of the peak. The resolution at 5.895~keV (the full absorption peak) is 17.5~$\pm$~0.1\%. The sum of the escape and fluorescence peaks have an energy resolution of 43.0~$\pm$~1.6\% at 2.57~keV, the theoretical weighted mean energy of these two peaks combined.

\begin{figure}[tp]
\begin{center}
\begin{tabular}{c}
	\includegraphics[width=0.8\textwidth]{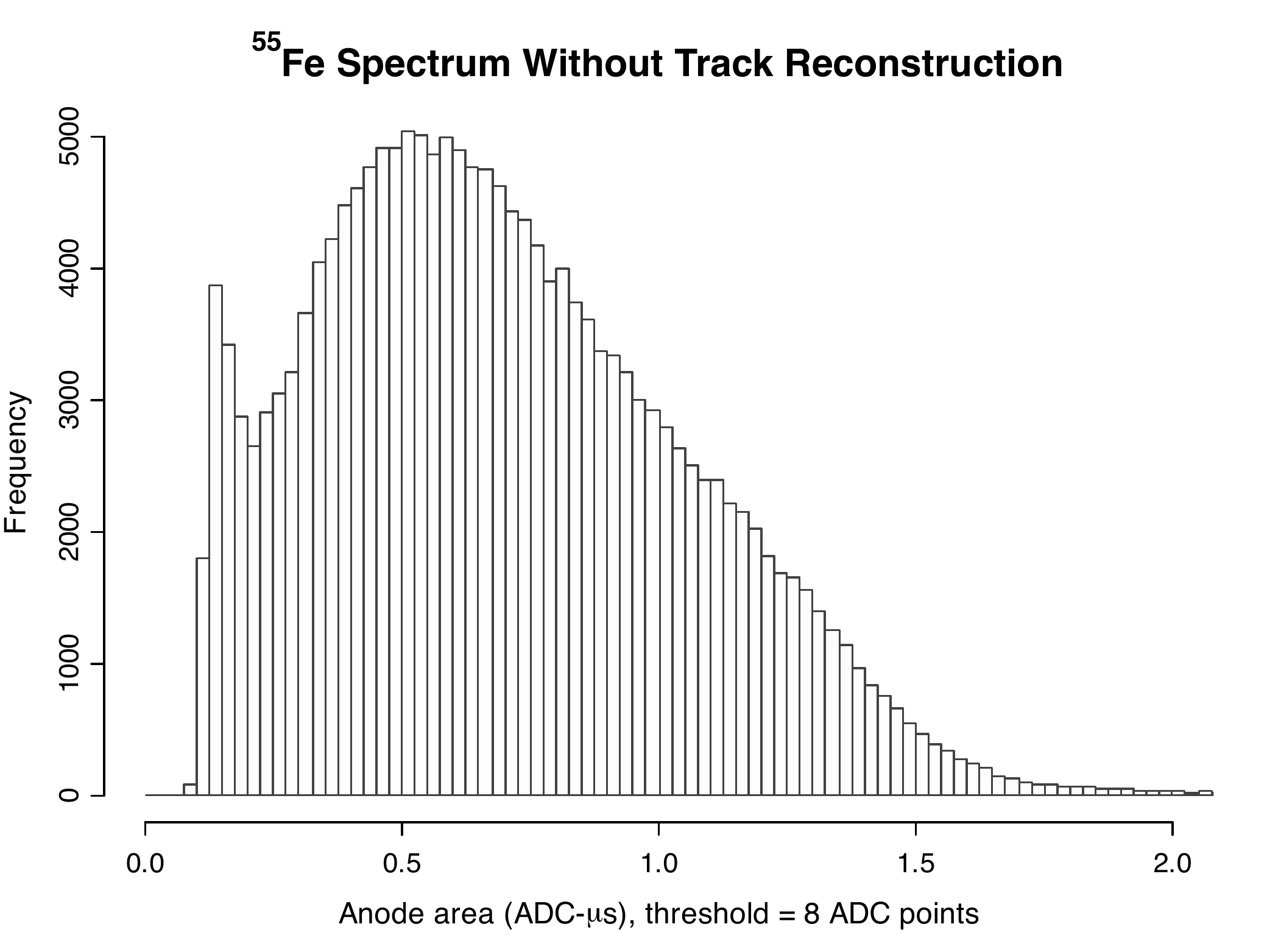}
\end{tabular}
\end{center}

\caption[]{The same \fe55 energy spectrum but where all events are treated as one-wire hits.}

\label{figure:sans-track-reconstruction}
\end{figure}

\begin{figure}[tp]
\begin{center}
\begin{tabular}{c}
	\includegraphics[width=0.8\textwidth]{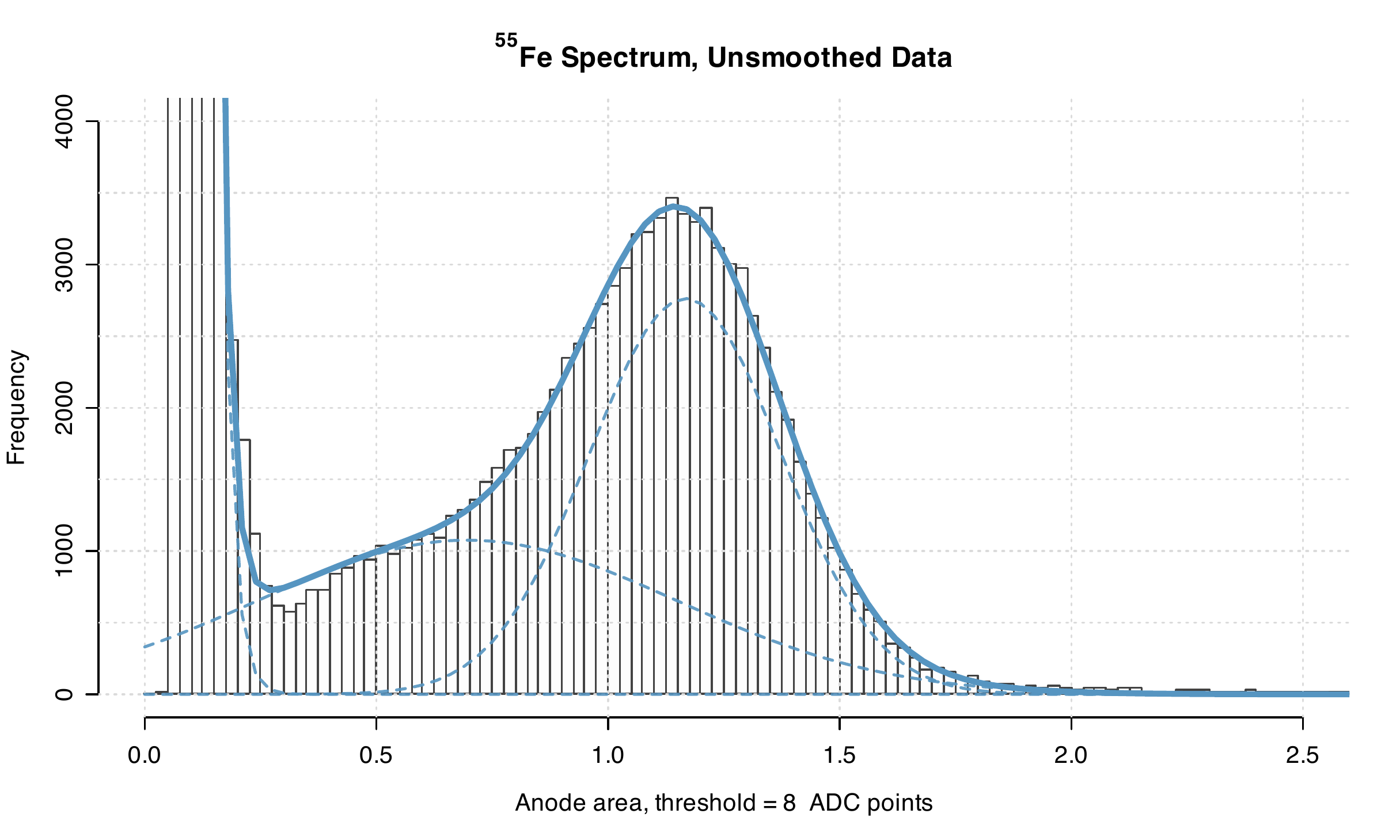}

\end{tabular}
\end{center}

\caption[]{The same \fe55 energy spectrum as presented in Figure~\ref{figure:full-spectrum} but where the Savitzky-Golay digital filter has not been applied. While the mean position of the full absorption peak has moved down by only 3\%, the energy resolution at the escape peak worsens from 17.5\% to 65.1\% and is no longer a clearly resolved feature.}

\label{figure:histogram-without-smoothing}
\end{figure}

Figure~\ref{figure:sans-track-reconstruction} shows the \fe55 energy spectrum when all of the individual pulses on the readouts are considered alone with no track reconstruction. The importance of identifying tracks that are divided on two wires is clear: while the events still show separation with the noise peak, a great excess appears at lower energies and a calibration cannot be reliably made. Additionally, the digital filter is shown to improve sensitivity at low energies. Figure~\ref{figure:histogram-without-smoothing} shows the energy resolution when the Savitzky-Golay filter has not been applied. Here the mean position of the full absorption peak has decreased slightly to 1.17~ADC-\microsec with a resolution of 17.8~$\pm$~1.0\%, and the mean position of the smaller peak is 0.696~$\pm$~0.087~ADC-\microsec and has a substantially higher energy resolution of 65.1~$\pm$~11.0\%.

\begin{table}[t]
\begin{center}
\begin{tabular}{lc}
\toprule
Source of Track & Energy (keV)\\
\midrule
Electron & 1.22\\
Alpha particle & 1.10\\
Carbon nuclear recoil & 1.93\\
Sulphur nuclear recoil & 2.90\\
\bottomrule

\end{tabular}

\caption[Potential DRIFT-II Energy Sensitivity]{Minimum potential detection energy based on the \fe55 analysis, using a value of 58 NIPs. The energy conversion is strongly dependent on the accuracy of the $W$ value, particularly at low energies.}
\label{table:fe55-potential-threshold}

\end{center}
\end{table}

\section{Potential Detector Sensitivity}
The energy spectrum of the \fe55 calibration provides an opportunity to characterise the potential energy threshold the \driftII{} detector. A reasonable value for the lowest energy that is distinctly above the noise peak from Figure~\ref{figure:full-spectrum} is 0.25~ADC-\microsec (defined here as the point where the extrapolation of the exponential distribution of the frequency of the noise falls below one). This is converted to a value of 58~NIPs given the mean position of the full absorption peak as 279~NIPs (for $W$~= 21.1~$\pm$~2.7~(stat)~$\pm$~3~(syst)~eV \cite{cs2-w-value}). The energy of an event that would produce this number of NIPs depends on the incident particle's $W$ value. Table~\ref{table:fe55-potential-threshold} lists several types of event and the energy that would generate 58~NIPs in 40~Torr \cstwo. The conversion from NIPs to keV was made by performing a linear interpolation between the origin and the smallest available value. $W$ values are taken from \cite{drift-neutron-recoils} (nuclear recoils) and \cite{2004NIMPA.516..406S} (alpha particles), although it should be noted that $W$ values are not well known below~20~keV \cite{majewski-head-tail}. Finally, an improvement can be made in energy sensitivity by increasing the resolution of the signal digitisation. Events at the lowest energies of a few keV typically measure 7--12~ADC counts in height, and this (easily corrected) quantisation creates a limit in the energy resolution.

Some caveats must be noted with these numbers. First, the thresholds presented here are indicative of the potential using the current technology and are not immediately applicable to a dark matter search. Running the detector with a trigger set at these values would generate too much data to reasonably analyse; for example, the values demonstrated here currently fall well within the peak-to-peak noise of the raw data. However, electronics are being designed to filter noise that is currently removed through software, and a more intelligent trigger can be designed to actively perform further cuts in near-real time. Thus, the detector could be made to run in a triggerless mode approaching or at these values. Second, the exact conversion depends on the accuracy of the $W$ values for each event type, and these have not been specifically measured for \cstwo. While some approximations are reasonable to use at higher energies, they become less valid at very low energies. Finally, track diffusion increases with drift distance, and thus the minimum event energy sensitivity will increase as a function of drift length. While the majority of the \fe55 events occur near the MWPC, the S fluorescence X-rays produced at these sites are produced isotropically and travel with a mean free path of 21~cm, as such, they represent a greater range of drift lengths that cover about half the maximum drift distance (25~cm). Events that occur further than this will undergo greater diffusion, and consequently the minimum detectable energy of any interaction will rise as a function of distance from the MWPC.

\section{Effect of Threshold Improvement on a WIMP Exclusion Limit}
Lowering the trigger threshold would have a positive effect on the ability for DRIFT-II to set a WIMP cross section limit. The current detector energy threshold is approximately 1250~NIPs, or $\sim$65~keV sulphur nuclear recoils. This is a soft threshold as the detector triggers on $dE/dt$ on the anode sum (AS) trace shown in Figure~\ref{figure:fe55-event}. This introduces a slight directional bias as tracks oriented along $dz$ (and thus $dt$) deposit energy over a longer period of time compared to the same track in a perpendicular orientation which will deposit all of its energy in a short amount of time.

\begin{figure}[tp]
\begin{center}
\begin{tabular}{c}
	\includegraphics[width=0.8\textwidth]{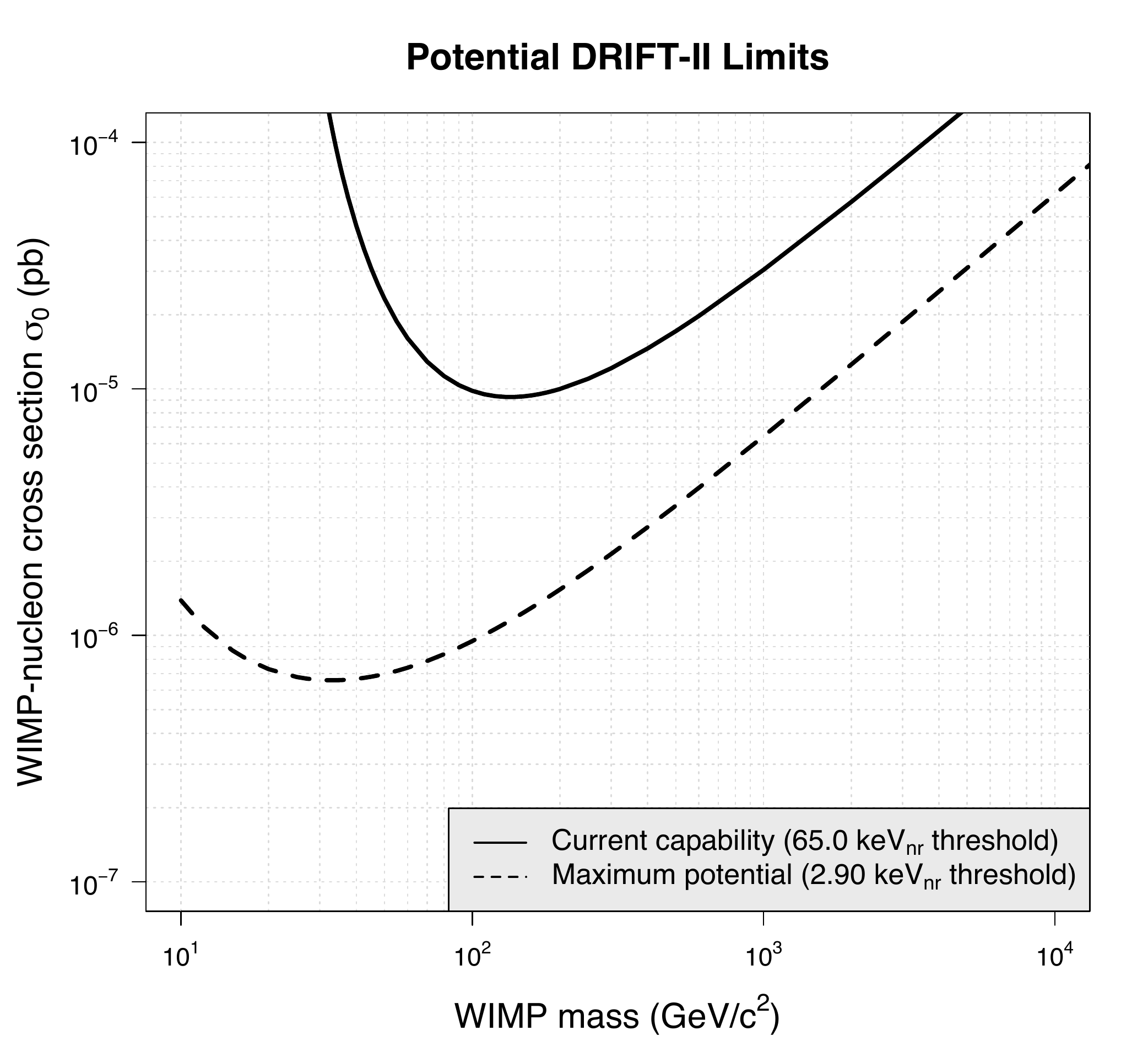}

\end{tabular}
\end{center}

\caption[Theoretical Exclusion Plot]{Two theoretical exclusion limits calculated to show the result of an improvement in energy sensitivity for DRIFT-II (see text for details and calculation inputs). Improvements to the detector have the potential to yield an experimental result that could lie between these two curves.}

\label{figure:limit-plot}
\end{figure}

Figure~\ref{figure:limit-plot} shows two theoretical limit calculations based on the following inputs: one year live time, zero background events, zero events detected, a target mass of 112.6~g (the mass of the sulphur in a single DRIFT module), an energy resolution of 17.5~$\pm$~0.1\%, and a constant detector efficiency of 80\%. The upper line shows the exclusion limit for DRIFT-II with an energy sensitivity down to 65~keV sulphur nuclear recoils, and the lower plot brings the sensitivity down to 2.90~keV. The result shows better than an order of magnitude improvement for lower WIMP masses (<~100~GeV/c$^{2}$), and roughly half an order of magnitude above that.

This figure is presented primarily to demonstrate the effect of the potential for the improvement in energy resolution using the technology of the currently operating detector. The detector's threshold cannot be set this low currently. Several factors will degrade the limit setting ability, but each can be ameliorated by varying degrees. A background of radon progeny events \cite{drift-alpha-paper} can be discriminated with the measurement of the absolute $z$ position in the chamber. New electronics are currently being tested that may alleviate the need to perform the smoothing techniques presented here and allow real-time operation with an energy threshold of a few keV. A gamma background may become a limiting factor at the lowest energies, but the discrimination capabilities can only be investigated once new electronics are in place. In practice and with improvements such as these, the actual ability of a single detector module will lie somewhere between the two curves in Figure~\ref{figure:limit-plot}.

\section{Conclusion}

This analysis demonstrates that the current operating technology of the \driftII{} detector using a 1~m$^{3}$ negative ion TPC with a gaseous \cstwo target has the potential to identify true particle interactions as low as 1--4~keV. Digital polynomial smoothing techniques are presented here to measure events of this scale from experimental data. As a track may hit several wires, it is not necessary to trigger on these small values, rather, the advantage comes in the ability to select small amounts of energy from individual wires to more accurately measure the total charge a track deposits on the MWPC\@. This improvement in energy sensitivity has the potential to improve the detector's limit setting capability by roughly an order of magnitude or better for low mass WIMPs and half that for higher masses. While reasonable improvements to the detector will need to be made to trigger at energies of a few keV in regular operation, significant progress in energy sensitivity is clearly achievable.

\section*{Acknowledgements}
The DRIFT collaboration would like to thank Cleveland Potash Ltd for continued support and use of facilities at Boulby Mine.


\bibliography{fe55_paper}

\end{document}